\documentclass[aps,prl,amsmath,amssymb,twocolumn,showpacs,superscriptaddress]{revtex4}
\usepackage{graphicx}

\begin{document}

\title{Diffusion-limited reactions in crowded environments 
}

\author{N.~Dorsaz}
\affiliation{University of Fribourg, Adolphe Merkle Institute, CH-1723 Marly 1, Switzerland }

\author{C.~De Michele}\affiliation{ Universita di Roma “La Sapienza”, Dipartimento di Fisica, P.le Aldo Moro 2, 00185 Roma, Italy}

\author{F.~Piazza}
\affiliation{ Ecole Polytechnique F\'ed\'erale de Lausanne (EPFL), Institute of
Theoretical Physics, Laboratory of Statistical Biophysics, 1015 Lausanne, Switzerland}

\author{P.~De~Los~Rios}
\affiliation{ Ecole Polytechnique F\'ed\'erale de Lausanne (EPFL), Institute of
Theoretical Physics, Laboratory of Statistical Biophysics, 1015 Lausanne, Switzerland}

\author{G.~Foffi}
\affiliation{Ecole Polytechnique F\'ed\'erale de Lausanne (EPFL), Institute of Theoretical Physics, 1015 Lausanne, Switzerland}

\begin{abstract}

Diffusion-limited reactions ($DLR$) are usually described within the Smoluchowski theory, which neglects interactions between the diffusing components. We propose a first extension of such framework that incorporates excluded-volume effects,  
considering  hard diffusing agents in the presence of an absorbing sink. 
For large values of the absorber-to-diffuser size ratio $R_s$, 
the encounter rate obtained from the simulations is in very good agreement with a simple generalization of the Smoluchowski equation at high densities. 
Reducing  $R_s$, the rate gets substantially depressed at intermediate packing and 
become even non-monotonic for $R_s \ll 1$.
Concurrently with the saturation of the rate, stationary density waves set 
in close to the absorbing center.
A mean-field, effective-potential analysis of the Smoluchowski equation under crowding
sheds light on the subtle link between such ordering and the 
slowing down of the encounter dynamics.
Finally, we show how an infinitesimal amount of non-reacting impurities 
can equally slow down dramatically the reaction.

\end{abstract}

\pacs{61.20.Lc, 64.70.Pf, 47.50.+d}

\maketitle


%


Processes such as gelation,  coagulation, crystallization or self-assembly in colloidal 
or polymer systems,  thin-film growth in materials science and chemical reactions in biology 
involve, as a first step, the random movement of particles suspended in a fluid~\cite{Family1984}.
The components diffuse and when they come into contact, a reaction may be triggered, such as aggregation or assembling. As an example, the  usual paradigm for biochemical reactions 
assumes the formation of an encounter complex, which may subsequently undergo chemical transformation, yielding the product.  In most cases, the fixation step proceeds much faster than the encounter, 
in which case the reaction is said to be {\em diffusion-limited}. Such reactions are commonly found in biochemical processes such as enzyme catalysis, protein aggregation and complexation in cells~\cite{Rice1985}. 

The simplest model of diffusion-limited encounter in three dimensions  has been formulated almost one century ago by 
Smoluchowski under the hypothesis of non-interacting, spherical and chemically
isotropic reactants, a diffusing particle (p) and a sink (s)~\cite{smoluch1917}, 
leading to the Smoluchowski absorption rate 
\begin{eqnarray}
\label{smoluchrate}
\kappa_{S} & = & 4 \pi D R \rho_{\infty}
\end{eqnarray}
where $\rho_{\infty}$ is the (relative) bulk density of the reactants, 
$D = D_s + D_p$ is the coefficient of relative diffusion and $R = R_s + R_p$ the
encounter distance. 

\begin{figure}[!]
\centering
\includegraphics[width=7.5cm]{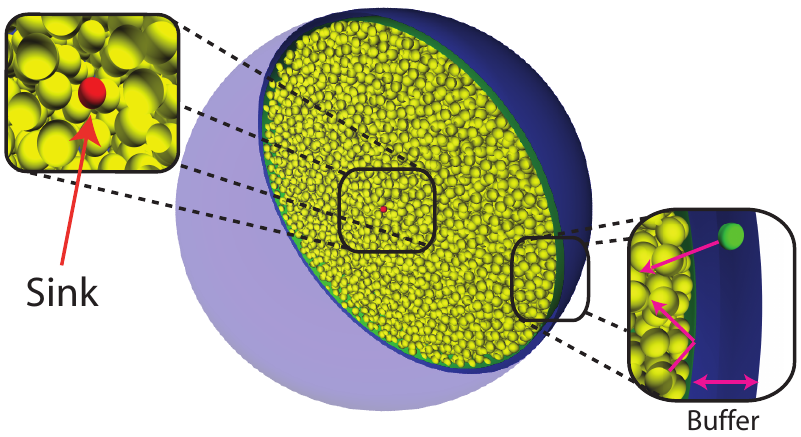}
\caption{ (Color online) Simulation scheme. The reacting
  particles (yellow spheres) diffuse in a spherical box. Upon reaching
  the absorbing sink (red sphere) located at the center of the box,
  they get absorbed and are reinserted in the external 
  buffer (green sphere).  }\label{F0}
\end{figure}

The Smoluchowski theory is still nowadays the main theoretical framework within which the 
aforementioned processes are analyzed.
However, this approach is strictly valid only for ideal, infinitely diluted solutions, while most systems become interesting at concentrations far from the ideal gas limit. Cells, for instance, contain a large number of proteins,  nucleic acids, and other smaller molecules that occupy up to $30$-$40\%$ of the available volume~\cite{Ellis2003} and that cannot overlap with each other. 
As a matter of fact, {\em crowding} 
effects are expected to impact profoundly on the thermodynamics and kinetics of biological processes in vivo~\cite{Ryan1988,Zimmerman1993,Ellis2003}, such as protein folding and stability~\cite{Cheung2005} and aggregation~\cite{Minton2000a}. Yet, the effects of crowding on diffusion-limited processes have been examined only  in the case of low density of diffusing particles~\cite{Dzubiella2005} or for tracers diffusing in a medium of inert particles~\cite{Dong1989,Sun2007,Schmit2009, Kim2010}.  
In colloidal science, instead, the properties of crowded environments have been investigated widely.  Experiments on concentrated suspensions of nearly hard-core particles revealed unexpected phenomena both in their thermodynamics and in the dynamics, spotlighting a subtle entwining between packing, structuration and dynamics~\cite{Pusey1986}.
\\
\indent In this paper, we generalize the classic Smoluchowski
problem to arbitrary crowding conditions by means of a novel computational scheme
that allows to efficiently explore the effects of increasing packing
on the encounter dynamics. Typical numerical simulations of diffusion-limited reactions
employ different declensions of the Brownian dynamics (BD) 
algorithm~\cite{Ermak1978,Cichocki1992}.  Here, we consider a liquid
composed of hard spheres of radius $R_p$ described through Event
Driven Brownian Dynamics ($EDBD$)~\cite{Foffi2005,Scala2007,DeMichele2010}.  This
technique allows to efficiently simulate the stochastic dynamics of
many-body hard-core systems. In order to simulate encounter reactions at finite
densities, we adapted the EDBD to the configuration of an absorbing
sink of radius $R_s$ located at the center of a spherical
bounding box with constant-flux boundary conditions (Fig.~\ref{F0})~\cite{Dorsaz2010}. 
%
%
In this way, by tuning the density in the box, we can 
measure the encounter rate $\kappa$ and the stationary density 
profile $\rho(r)$, from the infinite-dilution Smoluchowski limit, namely
\begin{equation}
\label{smoluchrho}
 \kappa = \kappa_{S},  \:\:\:\: \:\:\rho_S(r)=\rho_{\infty} \: [1- (R_s+R_p)/r]
\end{equation}
up to high crowding conditions. In the following, densities are expressed in terms of  
the corresponding packing fraction of the spheres $\phi$,  and lengths in units of $R_p$.
%
%
%
%

As a first investigation of crowding effects, we simulated the case 
of a large sink-to-particle diameter ratio ($R_s = 8/3$). At low densities ($\phi< 0.1$), 
the rate increases linearly, in agreement with the BD and analytical results by Dzubiella and
McCammon~\cite{Dzubiella2005}, following the first order virial correction to the Smoluchowski rate. Increasing the crowding further, the encounter rate is strongly enhanced, reaching even an 
eight-fold magnification at $\phi=0.4$ with respect to the infinitely diluted limit. 

In order to account for the strongly nonlinear increase of the rate at 
intermediate crowding, one can consider the Smoluchowski equation and
take inter-particle interactions into account in the expression of the net flux
$\overrightarrow{J}$ through a generalized Stokes-Einstein relation
for the collective diffusion coefficient~\cite{Dhont1996}. 
At first order, crowding effects can be accounted for through 
the density dependence of the pressure in the bulk liquid $\Pi(\rho)$ (vanishing at zero density), 
yielding a generalized diffusivity
$D(\rho) = D_0 \, d \, [ \beta \,\Pi(\rho)]/d\rho $, where 
$\beta^{-1} = k_{ B} T$
and $D_0\equiv\lim_{\rho\rightarrow 0}D(\rho)$.
%
The steady-state Smoluchowski equation in spherical coordinates reads then
\begin{equation}
\label{e:ssSmol}
\overrightarrow{\nabla}  \cdot \overrightarrow{J} = - \frac{\beta D_{0}}{r^2} \,
                                                \frac{d}{d r}
                                                \left\lbrace
                                                    r^2 \frac{d \rho(r)}{d r} \frac{d \Pi[\rho(r)]}{d\rho(r)}
                                                \right\rbrace = 0
\end{equation}
\noindent with the boundary conditions $\rho(R_s+R_p)=0$ and $\lim_{r\rightarrow\infty}\rho(r)=\rho_{\infty}$.
Integrating the equation that defines the encounter rate, namely 
$\kappa/(4 \pi r^2) = J(r)$, in the interval $[R,+\infty)$ with
$J(r) \equiv |\overrightarrow{J}| = \beta D_{0}(d \rho(r)/d r )(d \Pi(\rho)/d \rho)$ one obtains
%
\begin{equation}
\frac{\kappa}{\kappa_S} = \frac{\beta \Pi(\rho_\infty)}{\rho_\infty}
\label{rate}
\end{equation}
where $\beta \Pi(\rho_\infty)/\rho_\infty = Z(\phi)$ is the compressibility factor. The latter can be well  
approximated through the Carnahan-Starling (CS) equation of state  $Z(\phi)=(1+\phi+\phi^2 - \phi^3)/(1-\phi)^3$ ~\cite{Carnahan1969}.
At moderate densities, the virial expansion 
%
$\beta \Pi(\rho)=\rho(1 + B_2 \rho + O(\rho^2))$ yields the
Smoluchowski rate at zeroth order and the first-order correction
$\kappa/\kappa_{S}=1 + B_{2}\rho_{\infty}$ derived in Ref.~\onlinecite{Dzubiella2005}.

\begin{figure}[t!]
\centering
\includegraphics[width=7cm]{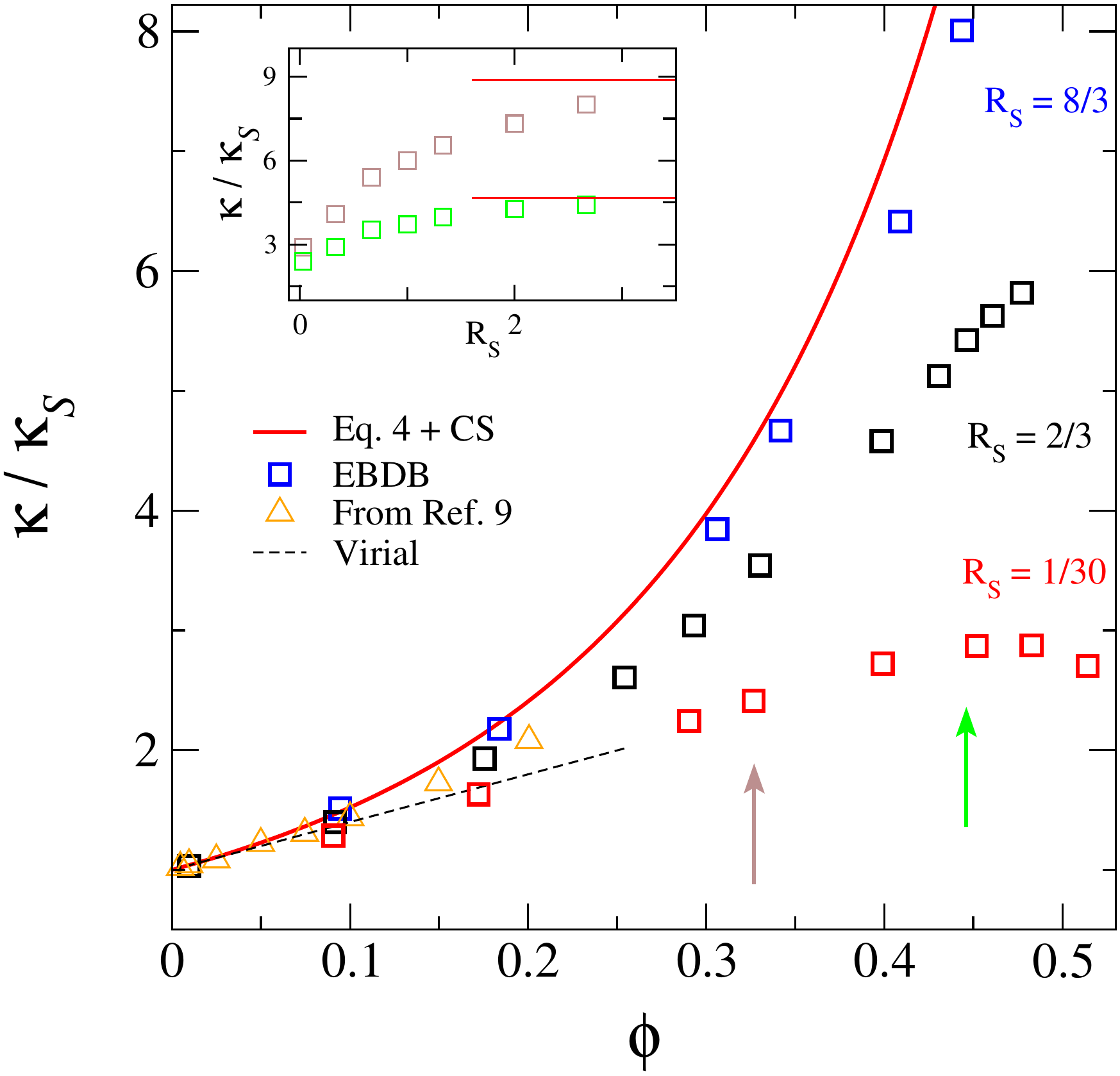}
\caption{ (Color online) Encounter rate versus crowding. 
For large sink-to-particle diameter ratios $(R_s=8/3)$, Eq.~\eqref{rate} accounts for the EDBD results. Reducing $R_s$, the rate saturates at high packing,
while it displays a maximum  for $R_s \ll 1$.  The virial solution and the BD results of Ref.~\onlinecite{Dzubiella2005} are also shown. In the inset, we plot the rates versus $R_s$ 
for $\phi = 0.33$ and $\phi=0.44$, together with the corresponding theoretical limits
predicted by Eq.~\eqref{rate}.}\label{F1}
\end{figure}

For large values of the sink size, the rate increases monotonically  
with the density in good agreement with Eq.~\eqref{rate}, 
which provides an accurate prediction in the limit $R_s \rightarrow  \infty$.
This is clearly illustrated by inspecting how the rate varies with $R_s$ at fixed
packing fraction (inset in Fig.~\ref{F1}). 

The density gradient between the sink and the bulk regions induces a net force towards the sink, 
the latter behaving as a depression in an high pressure environment. The rate is thus 
directly related to the difference in 
%
%
pressure between the diluted region near the sink and the denser bulk. 
Remarkably, however, when $R_s$ is decreased the rate gets substantially 
depressed at intermediate crowding and becomes even non-monotonic for $R_s \ll 1$,
featuring a maximum and a drop at high crowding. 

\begin{figure}[t]
\centering
2\includegraphics[width=7cm]{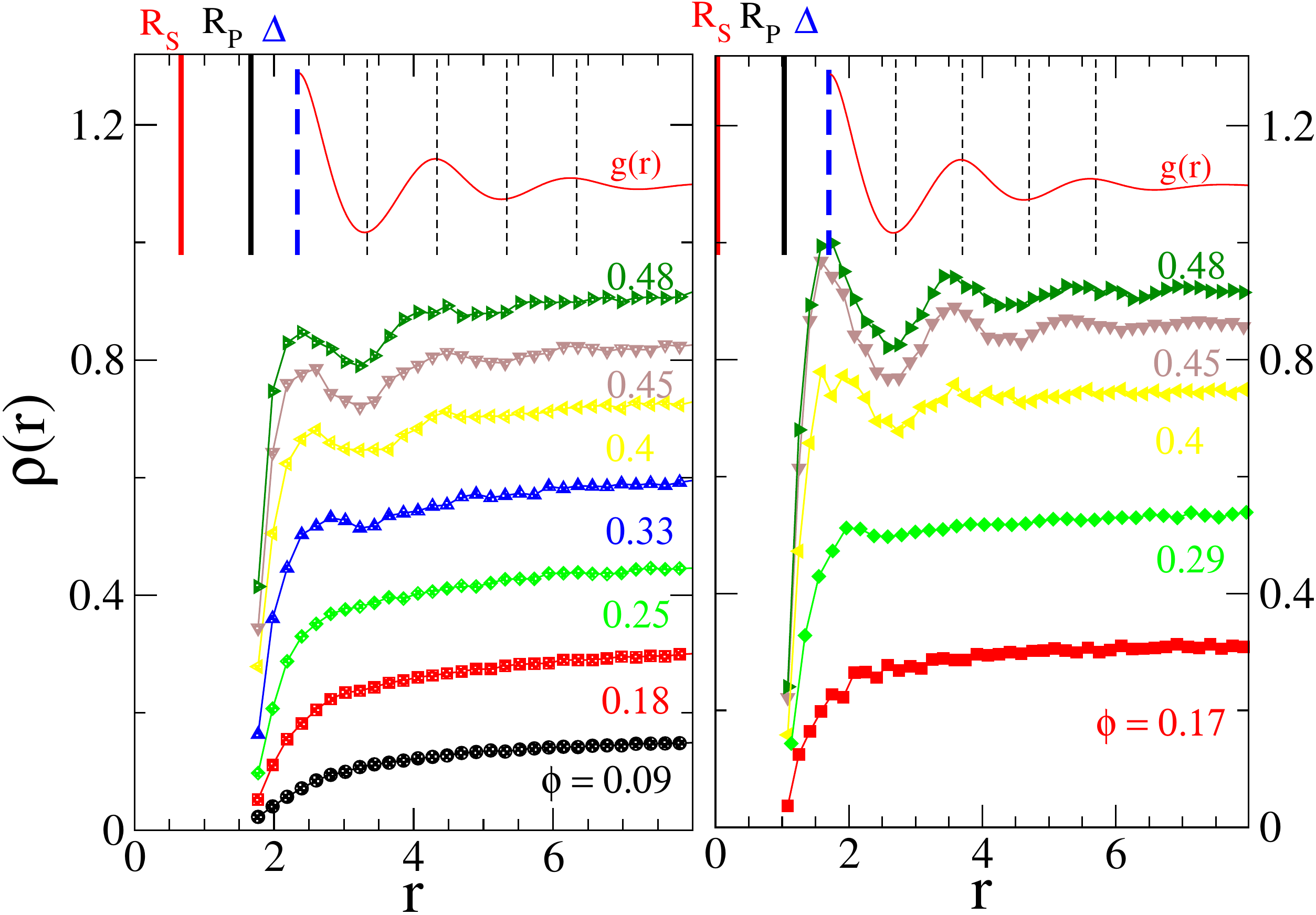}
\caption{ (Color online) 
Structuration of the density profiles under crowding. The profiles are plotted
at increasing bulk packing fraction $\phi$ for $R_s=2/3$ (left) and $R_s=1/30$ (right).
Structuration is enhanced in the case of the smaller sink, 
but the liquid gets structured at approximately the same distance $\Delta$ from the contact 
distance $R_s+R_p$ in the two cases. The density oscillations are in phase with the radial 
distribution function $g(r)$ of a hard-sphere liquid at intermediate densities.
}\label{F2}
\end{figure}

The stationary density distributions of the diffusing particles around the absorbing boundary surface 
offer a key to understand the drop in the rate observed at intermediate packing when reducing $R_s$. 
As illustrated in Fig.~\ref{F2}, we found that ({\em i}) oscillations set in progressively 
while increasing $\phi$ and ({\em ii}) 
the structuration is strongly enhanced for small values of $R_s$. 
More specifically, the accumulation of particles around the sink corresponding to the first maximum 
in the density profiles occurs at a distance $\Delta' \simeq (R_s + R_p) + \Delta$,  
where $\Delta$ does not depend on $R_s$.

As it is clearly shown in Fig.~\ref{F2}, the progressive structuration of the system 
bears the  clear signature of the radial distribution function $g(r)$ of a hard-sphere 
liquid at intermediate densities. 
These effect is by no mean trivial, as g(r) provides a measure of spatial correlations in the reference frame of a moving particle.
Here, translational symmetry is broken by the sink,  which causes the emergence of denser layer  near the absorber, triggering the onset of stationary density waves.

\begin{figure}[b]
\centering
\includegraphics[width=8cm]{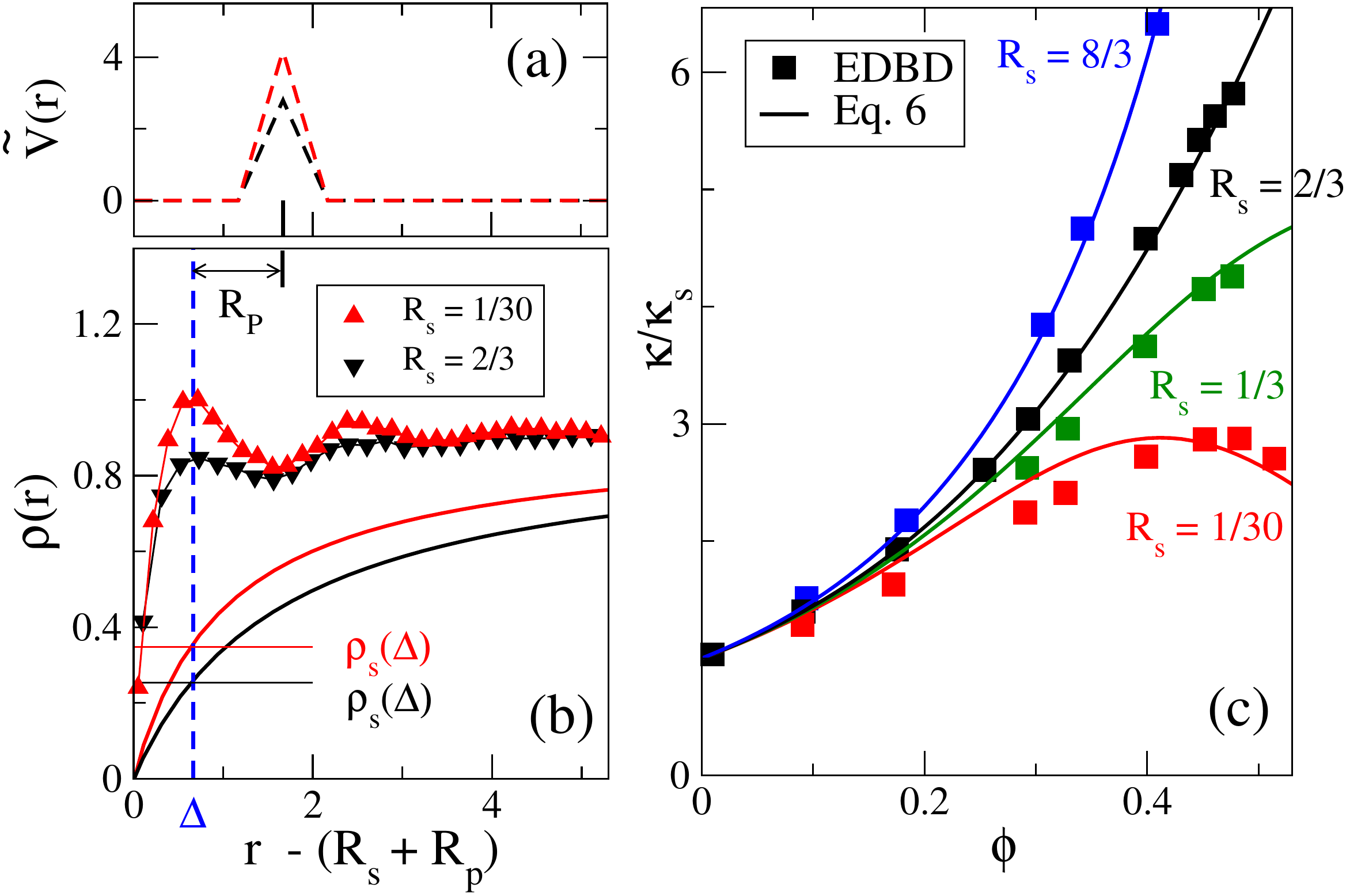}
\caption{(Color online) Effective potential description of $DLRs$ under crowding.  
The effective potential $V(\phi,R_s)$ 
(a) accounts for the accumulation of particles at distance 
$\Delta$, resulting in the oscillations in the density profiles (b).  
The solid lines in panel (b) are plots of the  Smoluchowski profiles~\eqref{smoluchrate}, 
emphasizing the $R_s$ dependence of the density at $\Delta$. 
\textbf{(c)} The one-parameter fits obtained through Eq.~\eqref{e:MFrate} 
are in excellent agreement with the numeric results for all values of $R_s$, 
providing a quantitative link between the drop in the rate and the 
onset of stationary density oscillations.}\label{F3}
\end{figure}

The simultaneous onset of density oscillations and rate falloff observed upon reducing $R_s$ 
suggests that the organized accumulation of particles 
is the consequence of a competition among the particles diffusing in the proximity of the sink. 
To this regard, it is important to note that, $\Delta$ being a constant, the  
particles compete for the sink at a distance from the absorber that depends on the size 
of the diffusers only, and not on the size of the latter.

In order to emphasize the connection between the increasing structuration 
of the liquid and the slowing down of the encounter process  
it is possible to regard the onset of density fluctuations as marking the 
appearance of an effective  potential between the sink and individual particles.
Within such representation, the accumulation of particles at $\Delta$ will act on a 
freely diffusing particle as an energy barrier $\tilde{V}(r)=f(r;\phi,R_s)$ (Fig.~\ref{F3} (a)),
whose height should depend parametrically on both the bulk 
concentration $\phi$ and the sink size $R_s$ 
to account for the increasing structuration of the density profile for large 
values of $\phi$ and small values of $R_s$. 
In general, in the presence of an interaction potential $V(r)$,
the Smoluchowski rate reads~\cite{Houston1985},
\begin{equation}
\label{e:kspot}
\frac{\kappa}{\kappa_S} = \left[  
                             R \int_{R}^\infty \frac{e^{\beta V(r)}}{r^2} \, dr
                          \right]^{-1}
\end{equation}
It is possible to generalize Eq.~\eqref{e:kspot} in order to account for 
the effects of crowding by noting that finite densities imply 
$\kappa \propto \beta \Pi(\rho_\infty)$, as shown by Eq.~\eqref{rate}, instead 
of $\kappa \propto \rho_\infty$, as from Eq.~\eqref{smoluchrate} (free particles) 
and Eq.~\eqref{e:kspot} (sink-particle potential). As a consequence, we may write
\begin{equation}
\label{e:MFrate}
\frac{\kappa}{\kappa_{S}} = \frac{\beta \Pi(\rho_\infty)}{\rho_\infty}   \left[  
                             R \int_{R}^\infty \frac{e^{\beta \tilde{V}(r)}}{r^2} \, dr
                          \right]^{-1}   
\end{equation}.
It is reasonable to assume that the height $V$ of the potential energy barrier 
reflects the excess pressure that accumulates in the vicinity of the sink 
with respect to the ideal case 
as the packing fraction increases. As a consequence, we assume
\begin{equation}
\label{e:barrier}
\beta \tilde{V} = c \, \frac{\Pi(\eta) - \Pi_{id}(\eta)}{\Pi_{id}(\eta)} = c [Z(\eta)-1]
\end{equation} 
%
where $\eta$ is related to the local density at a distance $\Delta$ (Fig.~\ref{F3} (b)) and $\Pi_{id}(\rho)=\beta\rho$ is the ideal gas pressure.
For simplicity, we take $\eta =\rho_S(\Delta)$ which introduces the dependence on $R_s$ in the effective potential.


Remarkably, a one-parameter fit of Eq.~\eqref{e:MFrate} to the numerical data accounts extremely well for both the $\phi$ and $R_s$ dependence of the rate (Fig.~\ref{F3} c)  with a single best-fit value of 
the floating parameter ($c=2.1$) for $0<\phi<0.52$ and $1/30<R_s<8/3$, 
thus substantiating our theoretical analysis.
Despite its simplicity, this approach seems to capture the essence of the problem, namely a density-weighted competition among the particles diffusing around the sink. Diffusion-limited reactions in  crowded environments are thus the result of a subtle combination of the liquid properties of the diffusers with 
the underlying long-range Smoluchowski description.

\begin{figure}[]
\centering
\includegraphics[width=6.8cm]{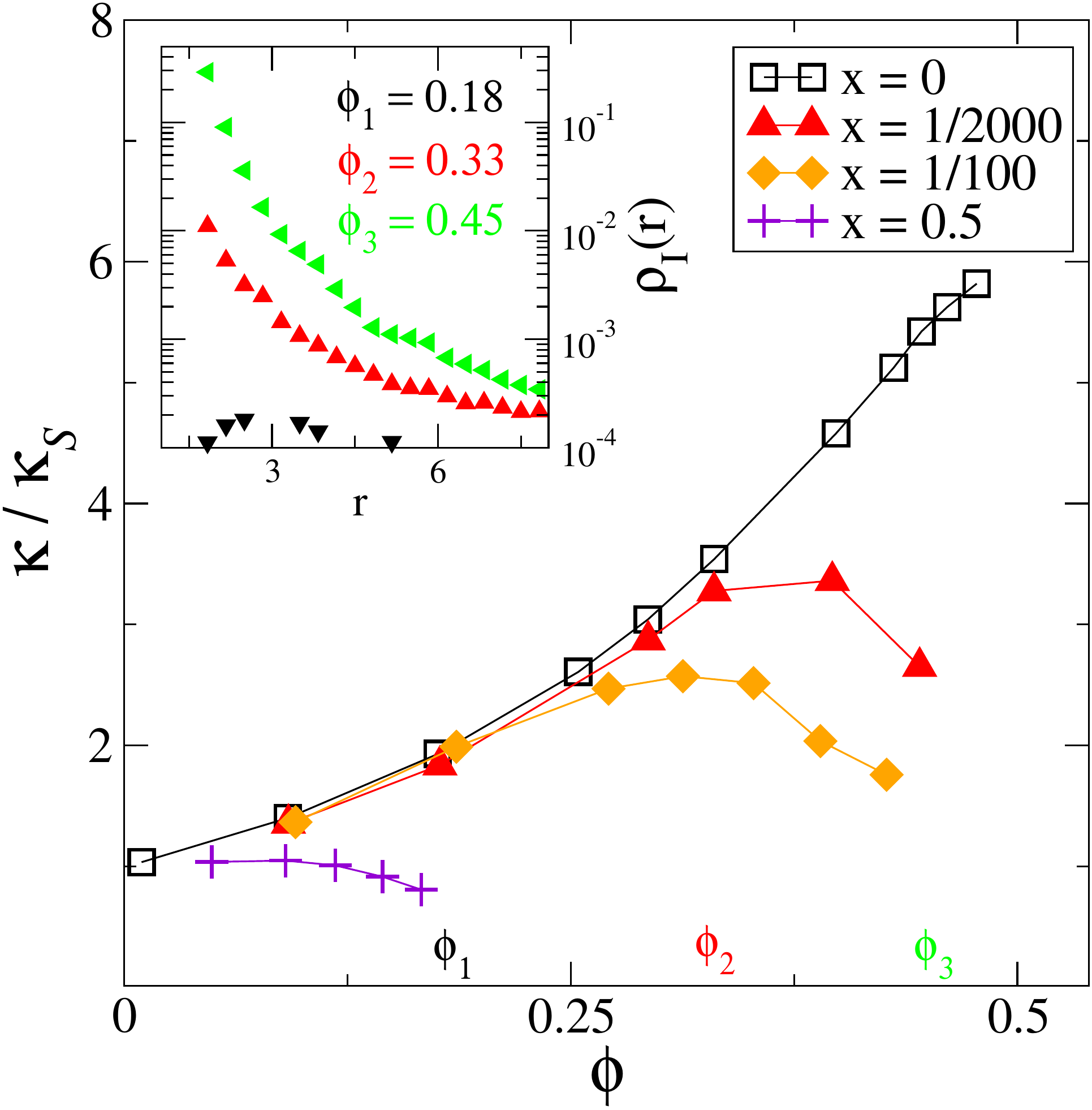}
\caption{(Color online) Encounter rate versus fraction of non-reacting impurities $x$ for $R_s=2/3$. 
Inset: the  density profile for impurities $\phi_I(r)$ for $x=1/2000$.}\label{F4}
\end{figure}

As a first extension of our numerical scheme to  
more complicated diffusion-limited reactions~\cite{Minton2000a,Ellis2003,Zaccone2010}, 
we investigated the effects of introducing a finite amount of non-reacting hard sphere \textit{impurities} 
($I$) in the system. 
We found that an infinitesimal fraction ($x=1/2000$) is sufficient to slow down considerably the 
reaction with respect to the one-component case (Fig.~\ref{F4}). While at low crowding the rate and the density profiles are insensitive to the presence of non-reacting particles, an accumulation of the 
latter against the sink takes place when increasing $\phi$ above 0.35. As a result, the sink 
gets screened to the reacting particles and the rate falls off. 
Indeed, when $\phi$ is increased, the impurities get trapped at the sink surface because 
of the large density gradient, and need to overcome a substantial energy barrier in order to 
get back to the bulk region~\cite{PAPER}.
The extreme sensitivity of the reaction rate to infinitesimal amounts of impurities might 
for instance be of great interest in the context of molecular trafficking across the nuclear 
pore complex, a protein complex responsible for protected exchange  
between the nucleus and the cytoplasm, or for preventing the transport of  
undesirable material through the nuclear envelope~\cite{Alber2007}.
\\
\indent In conclusion, in this letter we tackled a fundamental open issue. Our novel 
approach clearly highlights the physical ingredients necessary for the derivation 
of a self-consistent Smoluchowski theory valid for crowded environments. Moreover, 
while jamming phenomena like arches at bottleneck or at escape issues are well known in 
granular media or traffic studies~\cite{Helbing2000}, this is, to the best of our knowledge, 
the first time that a complex ordering in the presence of an absorbing boundary is reported 
for a system that is still in a liquid-like state. 
\\
\indent  ND acknowledges support from the Adolphe Merkle Foundation,  CDM by ERC (226207-PATCHYCOLLOIDS), SNSF ( visiting grant IZK022-
121268 ). GF acknowledges support by  the SNSF (grant PP0022\_119006).


\begin{thebibliography}{}
%

\bibitem{Family1984}
F. Family, D.P. Landau
\newblock Kinetics of Aggregation and Gelation
\newblock  ( North-Holland, The Netherlands, 1984).



\bibitem{Rice1985}
S.A. Rice,
\newblock {\it Diffusion-Limited Reactions}.
\newblock (Elsevier, 1985).



\bibitem{smoluch1917}
M.~V. Smoluchowski,
\newblock {\it Z. Phys. Chem.} \textbf{ 92}, 129 (1916).



\bibitem{Ellis2003}
R.J. Ellis, A.P. Minton,
\newblock {\it Nature} \textbf{425}, 27 (2003).

\bibitem{Ryan1988}
T.A. Ryan et al,
\newblock {\it Science} \textbf{239}, 61 (1988).


\bibitem{Zimmerman1993}
S.B. Zimmerman, A.P. Minton,
\newblock {\it Annu Rev Biophys Biomol Struct} \textbf{22}, 27 (1993).





\bibitem{Cheung2005}
M.~S. Cheung, D. Klimov, D. Thirumalai,
\newblock {\it Proc Natl Acad Sci } \textbf{102}, 4753 (2005).


\bibitem{Minton2000a}
A.~P. Minton,
\newblock {\it Curr Opin Struct Biol}, \textbf{10}, 34 (2000).







\bibitem{Dzubiella2005}
J.~Dzubiella, J.~A. McCammon,
\newblock {\it J. Chem. Phys.} \textbf{122}, 184902 (2005).

\bibitem{Dong1989}
W.~Dong, F.~Baros, J.~C. Andre,
\newblock {\it J. Chem. Phys.} \textbf{91}, 4643 (1989).

\bibitem{Sun2007}
J. Sun, H. Weinstein,
\newblock {\it J. Chem. Phys.}  \textbf{127}, 155105 (2007).

\bibitem{Schmit2009}
J.D. Schmit, E. Kamber, J. Kondev,
\newblock {\it Phys. Rev. Lett.} \textbf{102}, 218302 (2009).

\bibitem{Kim2010}
J.S. Kim,  A. Yethiraj,
\newblock {\it Biophys. J.} \textbf{98}, 951 (2010).



\bibitem{Pusey1986}
P.N. Pusey, W. van Megen,
\newblock {\it Nature} \textbf{320}, 340 (1986).




\bibitem{Ermak1978}
D.L. Ermak, J.A. McCammon,
\newblock {\it J. Chem. Phys.} \textbf{69}, 1352 (1978).

\bibitem{Cichocki1992}
B.~Cichocki, K.~Hinsen,
\newblock {\it Phys. A} \textbf{187}, 133 (1992).


\bibitem{Foffi2005}
G. Foffi et al.,
\newblock {\it Phys. Rev. Lett.} \textbf{94}, 078301 (2005).

\bibitem{Scala2007}
A. Scala, T. Voigtmann, C. De Michele,
\newblock {\it J. Chem. Phys.} \textbf{126}, 134109 (2007).

\bibitem{DeMichele2010}
C. De Michele
\newblock{\it J. Comput. Phys.}  \textbf{229}, 3276 (2010).

\bibitem{Dorsaz2010}
N. Dorsaz et al.
\newblock {\it J. Phys. Cond. Mat.} \textbf{22}, 104116 (2010).


\bibitem{Dhont1996}
J. Dhont,
\newblock {\it An Introduction to Dynamics of Colloids}.
\newblock (Elsevier, Amsterdam, 1996).


\bibitem{Carnahan1969}
N. Carnahan, K. Starling,
\newblock {\it J. Chem. Phys.} 51, 635 (1969).

\bibitem{Houston1985}
P.~L. Houston.
\newblock {\it Chemical Kinetics and Reaction Dynamics}.
\newblock Dover Publications, 1985.




\bibitem{Zaccone2010}
A. Zaccone et al.
\newblock {\it Phys. Rev. E} \textbf{80}, 051404 (2010).


\bibitem{PAPER}
\newblock{Details of the simulations reported in a forthcoming study.}


\bibitem{Alber2007}
F. Alber et al.,
\newblock {\it Nature}  \textbf{450},  695 (2007)

\bibitem{Helbing2000}
D. Helbing, I. Farkas, T. Vicsek,
\newblock {\it Nature} \textbf{407}, 487 (2000).


\end{thebibliography}


\bibliographystyle{apsrev}


\end{document}